\documentclass[a4paper,10pt]{article}

\usepackage{amsmath,amssymb,graphicx}
\usepackage{eurosym}
\usepackage[margin=2cm]{geometry}
\usepackage{array}
\usepackage{color}

\title{{\sc RFC-7800 -- Money Over Ip}\\{\tiny version $\alpha{}$3}}

\author{Laurent {\sc Fournier} -- {\small \{laurent.fournier@pluggle.fr, laurent.fournier@cupfoundation.net\}}\footnote{CEO {\em Pluggle} SAS - Toulouse - France. {\sc CupFoundation} is a non-profit association for promoting the $\sqcup$ currency.} }

\begin{document}
\maketitle

\begin{abstract}This {\sc Request For Comment} (RFC) 
 is a proposal for a new protocol to use {\bf Money Over the Internet protocol} ({\sc moi}). Features like a distributed architecture, a published cryptographic algorithm, a minimal authority responsibility, the absence of fees will make this protocol a perfect tool for citizens in today's digital World.
 
An implementation has validated the main principles and we entering now a testing phase (v0.1). Depending of the results, a released date for the 1.0 revision will to decided to allow anybody to send or to receive money to/from anyone, in any currency, with a regular and personal smart-phone. A distributed hash table ({\sc dht}) is used to store all transactions, public keys and certificates redundantly on several nodes. The all system may replace coins, banknotes and classical checks in the future. We also argue that the {\em Bitcoin} technology does not satisfy the requirements for a digital mean of payment. This proposal is expected to be reviewed and commented by the {\sc Internet Engineering Task Force}.
\end{abstract}

\section{Introduction}
There always has been two ways money can be used between citizens of the same community.
In the first and classical way, an authority is in charge of forging tangible coins and paper banknotes to facilitate trading, exchange of goods and services. That authority is usually unique to have such privilege.

The other way is to chose a unique trusted accountant who should record on a big blackboard or a big book all transactions between citizens and should allow a limited fare credit for each one. Nothing can be deleted on the board. Anyone can see all transactions to compute the balance of anybody.

If the second solution have been avoided in History for obvious reasons, making it impracticable for a large community, things has now to be reconsidered with three major innovations:

\begin{itemize}
\item {\bf Internet} as the Net of all nets connects every one on Earth. 
\item {\bf Cryptography} advance techniques, and {\it elliptic curves} cryptography ({\sc ecc}) for highly secure transactions, authentication, certificates.
\item {\bf Peer2Peer} protocols making distributed/redundant shared data possible and not deletable.
\end{itemize}

We would like to show with the following RFC proposal that "{\sc Money Over Ip}"\footnote{we may use the acronym `{\sc moi}' for our proposal system. It means `me' in French, to enforce the concern of individuals for money.} is not only possible, but also simple, efficient and compliant to the main democratic requirements for the money function.

Our main goal is to provide a free of charge payment system (not like {\em Visa/Mastercard/PayPal !}) at digital age, just like the usage of coins and paper banknotes has been free of charge before the Internet.

{\sc moi} is not a {\em cryptocurrency} in a sens that new money is never created from nothing. Every {\sc moi} account has to reference the same amount of money from an official banking account (\euro{}, \$, \...) and is using the same currency unit. However, it is based on advanced open-source cryptographic primitive to manage transactions securely and with a full transparency on Internet. This is more a {\bf crypto-payment} system.

\section{Digital check}
A pillar of the proposed system is the notion of {\em digital check}. What is a check in the physical World? It is a piece of paper with the following fields:

\begin{enumerate}
\item the current date,
\item the name/symbol of the currency used,
\item the amount of money to be exchanged,
\item the sender's name,
\item the recipient's name,
\item optionally a reference of an invoice
\item ...and the handwritten {\em signature} of the sender.
\end{enumerate}

In the digital World, many types of digital signatures exists nowadays. Let us select one of the shortest; EC-DSA with a specific curve called P-521 from the NIST\cite{nist}. Such a signature is always 132 long binary array for any message to be signed. In fact a {\it hash} of the message is used, so the signature itself does not have to worry about the length of the original message. What is important here is that any bit modification, for instance the amount of money, or the recipient id, will make the signature verification fail. Also, the message and the signature are never encrypted, they are plain readable text (if extended from binary) and anyone can verify whether the signature is valid or not on a given message.

Our proposal for the format of a transaction is the following: 

\begin{enumerate}
\item 1 byte: protocol version + reference field length (see bellow),
\item 3 bytes: the amount of money (cents for \$ or \euro) so the maximum transaction is around 167.000 units~!
\item 4 bytes: currency + date of the transaction
\item 8 bytes: the ID of the sender,
\item 8 bytes: the ID of the recipient,
\item 0 to 64 bytes: the optional reference (free signed text),
\item 132 bytes: the signature of the previous 24 bytes (88 bytes maximum) message by the sender.
\end{enumerate}

This first four bits of the first byte have the following value:

\begin{itemize}
\item 0x0: {\sc moi} protocol v$\alpha{}$ (test only)
\item 0x1: {\sc moi} protocol v0.1 (test only)
\item 0x2: {\sc moi} protocol v0.2 (test only)
\item 0x3: {\sc moi} protocol v0.3 (test only) 
\item 0x4: {\sc moi} protocol v1.0 (validated for trading)
\item 0x5: {\sc moi} protocol v1.1 (validated for trading)
\item 0x6 - 0xF Spare for {\sc moi} protocol v1.X 
\end{itemize}

The last four bites of the first byte indicates the length (in 4 bytes words) of an additional reference field.
Value zero means "no field" and the maximum value 16 (0xF) means 64 bytes added for a free text.
Be aware that the reference field is included in the signed message, so it can not be changed.
This is useful to give a reference of the exchanged good, of an invoice or any short message. 

\subsection{Date format}
The date shall have a specific format to prevent abusive check publishing; a transaction with a date in the future won't be accepted. 
Dates shall start the first January 2016 with a resolution of one minute. We could have second or less as resolution but this low frequency is to prevent high frequency financial transactions and more generally robot money spending. For any human-being, buying something every minute is far enough.
Date shall be coded on 26 bits so it ends in year 2143!

\subsection{Currency format}
The currency is coded in 6 bits, allowing 64 different currencies. Her some reserved values
\begin{itemize}
\item 0x00: "none" (test only)
\item 0x01: \$ [{\sc usd}]
\item 0x02: \euro [{\sc eur}]
\item 0x03: \pounds [{\sc gbp}]
\item 0x04: [{\sc cny}]
\item 0x05: [{\sc jpy}]
\item 0x3F: $\sqcup$ \cite{lfournier2}
\end{itemize}

So a transaction is {\bf 156 bytes} (220 bytes maximum) long in binary format and 195 characters (275 characters maximum) long with a {\sc base85} readable encoding.
Other representations can be used in ASN1, XML, JSON, but it is important to keep the transaction format as simple and as small as possible.

Thus, a transaction can be sent by e-mail, socket, {\sc http} request or with two {\sc sms}:
\begin{itemize}
\item one {\sc sms} including the message and a transaction-reference-id
\item the other {\sc sms} with the signature only and the same transaction-reference-id to recompose the full transaction
\end{itemize}
The transaction-reference-id can be 8 bytes long to fit in a 140 bytes {\sc sms}.

No-one can deny having written a valid digital check. As soon as a digital check is published on the Net and is valid, the transaction is considered executed.

For convenient, User ID are just the 8 ending bytes of the public key (132 bytes long), assuming we are using an asymmetrical cryptographic scheme for signature (DSA).

To start using the money over the Net, any user must have a mobile device, usually a smart-phone. At initialisation, it is recommended to disconnect the device from the Net. The user is invited to launch a random generator that produces several {\bf public-key/private-key} couples. She has to select one couple exactly as a young person chose a hand-written signature for all her life. No authority, no Secret Agency can be present to see/force someone to chose that couple and not that one. The private key is left private inside the phone and only on this device. We will see later the recovering process when a phone is stolen/lost/broken.
As one may expect, the public-key is published on the Web, as for any upcoming transaction.

The recipient of a check have to verify the signature validity but also that the public-key of the sender has been previously published and lastly that the account of the sender will have a positive balance if the transaction is accepted.

But how are balances values computed over the Net ?

\section{Get balances}
As a master rule, all accounts start with a null balance, and anytime after, the sum of all balances are always null, by design.

This mean that a few accounts are allowed to have {\bf negatives} balances, and we shall call them {\sc i-banks}. The other accounts are not allowed to have negative balances. Those are just regular accounts, for citizens.

An {\sc i-banks} account has some additional features so that nobody can decide by its own to become an {\sc i-bank}. In fact, {\sc i-bank} accounts are certified, unlike regular accounts. What does is mean exactly ?

In our system, it shall exist a {\bf master authority} ({\sc ma}), represented by a set of individuals so any decisions of that authority needs the signature of all the members. The master authority gives to a trust-able institution the right to have during a given period (one year default), a negative balance of a bounded value (may be 1B\euro{} for instance). The so called {\sc i-bank} receives a digital certificate, that is mainly her identity message signed by the master authority.

The master authority public key and all {\sc i-banks} public-key are all published on the Net, like regular accounts. Also all certificates are published. Thus anyone on Earth, connected to the Net can verify any transaction/certificate signatures but also that all transactions issued by {\sc i-banks} results balances that are compliant to the contract given by the certificate. 

An {\sc i-bank} is usually a well known institution that will always write a negative transaction on its Internet account by doing the symmetrical operation on the private banking account of its customer. For the end user, it's just like transferring some money from the bank to its Internet account.

Now, to compute a balance of a given account, it is straightforward. Just sum all the transactions, positive and negative, where the user acts as a sender or as a recipient. Anyone can compute the current balance of anyone.

Transactions do not have to be chained like for {\em Bitcoin}. Nodes can sort them by date and compute a small check-sum for each user. It'is then easy for each node to compare check-sums with other nodes and to know if it needs updating. However, the synchronization mechanism between nodes may produce the situation where two nodes have accepted two different transactions that leave positives balances at time of publishing, but when added all transactions, the balance become negative for the last transaction. Then the account is someway blocked until it receives a credit to be back positive. As a date is included in the transaction message, nodes may uses a tempo function of the check age to valid the transaction, so the risk that two transactions validated on two nodes at the very same time is very improbable. 

Because there may be thousands of transactions, balances are always computed and verified in a background process and the result is intermediately stored until a modification occurs. This is just a trick to get balance result in real time, but the principle is always to sum transactions values from the beginning. 

Servers that store data (public keys, transactions, certificates) are distributed and redundant, so they always compete each other to detect if another server has accepted a not fully valid transaction, in order to black list it. 
Not all servers should have a copy of wild-world transactions. A clever cache mechanism makes possible for a server to have only a small subset of users/{\sc i-banks} accounts just because there are only transactions inside such subset.

Server are distributed likes nodes of the {\em BitTorrent} protocol. However, a user is sharing a part of her CPU and disk space for {\sc money over ip} protocol, but she has no control on the transactions. Her computer might store their own transactions and also transactions from other ones, not especially friends.

Consider a classical high definition movie of 4GBytes stored on a basic PC hard disk. The same size can store about twenty five millions of transactions !
There is not rule or imposed technology to make persistent all the data. However we recommend to use a {\em key-value} database to be able to find instantly a transaction, a public key or a certificate.

The proposed format for certificate is the following:
\begin{itemize}
\item 1 byte: currency + protocol version (no reference field),
\item 3 bytes: the authorized {\bf debt} in the given currency in kilo (cross 1000);  maximum 16 B\euro{} for instance,
\item 4 bytes: the deadline of the certificate validity, 
\item 8 bytes: the ID of the i-bank account,
\item 132 bytes: the signature of previous message by the {\em master authority}. This signature is the resultant of several individual signatures.
\end{itemize}

A certificate is then 148 bytes long.

The {\sc ma} id and its public key is hardly written in the protocol.
For v0.1 testing protocol the {\sc ma} is defined by:
\begin{itemize}
\item  {\sc ma} ID (base85): \verb!Lfeeeb)XsP! or \verb!0x42DB8CD275A019E9!

\item  {\sc ma} Public-key (base85): \begin{verbatim}
0o;@Dnke406Ks)?ZJ}hg>~!A1ceYz@3V5;|u{w}LYCcPrxRm7@j0R%M(-P6avd@#8U)#;=iUp}S{~|GrA-4bk
0o+RoE>hbz;Yrc0dCjBO(on?P2_B12G8p*NLr4m8Kb%i&dWn(cuf)}Ry<+O{bmQOjQ;!fU(o{m*jM8<W8R-B3
\end{verbatim}

\item  {\sc ma} Public-key (base64):
\end{itemize}
\begin{verbatim}
AdyT8Joo3rMTbJGQbaFktux03sB3tltuCniyHbE6kENqPkuWuJTlGowGYsvTEtHBss-TnF_bzeuKBak4_yIxjCG3
AdxLCS5S2zbhSdGtec2j19JQxNsJHotPMhj400NICnI_nE9seomR5a_E1Xu9YurwdOPf9FOPECvSVELbjNJ1oBnp
\end{verbatim}

\section{Clearing house}
As soon as it exists more than one {\sc i-banks}, a clearing house shall be needed both for both the legal currency an for the {\sc moi} accounts. This is obviously a full automated task on the Internet. The {\sc ma} shall delegate an organization to work as a world-wide global clearing house. Each recorded transaction shall have a bit to remind if the transaction is cleared or not. 

\section{Recovering process}
Using asymmetrical keys for signing gives us the power to define our own digital identity. But that identify must be keep safe by its owner. That's why we recommend a three criteria authentication before any usage of the private key for signing.

\begin{itemize}
\item a mobile device you own and you can carry out anywhere,
\item some secret you know like a pass-phrase and you never share it,
\item some feature on you like your fingerprint ({\em TouchID iPhone}),
\end{itemize}

But what append if you forget your pass-phrase or if your phone is stolen/broken or lost ?
Well, it is always possible to buy another phone and reset the process, but what about the money on the last account ?

If you loose a phone without fingerprint authentication and with the pass-phrase hand written on top of it, it will be hard for any system to insure your money back...this is like dropping a banknote on the street.

There is however a simple way to protect the money on an Internet account. Just prepare (after initialisation) and sign one or several digital checks for an amount higher than your regular balance and select as recipient someone you trust in, a parent or friend, but {\bf do not publish} the check over the Net. Keep it on a saved usb-key or print it (200 characters) and save it just like a bank-note under your bed.

If for instance you have lost your phone with a 976\euro{} balance (you can read the exact balance value on any computer over the net, with your id that is part of your public key), and you had signed and recovered a check of 1000\euro{} somewhere. Put some money on your new account, send a check of exactly 24\euro{} to the stolen account and just after, publish the recovery check of 1000\euro{} from the old account to your friend.
Then, if the thief has not been as responsive as you are, the old account balance is now null and you just have to call your friend and ask her to refund the 1000\euro{} to your new account.
If you did not prepare a recovery check and if the account is not broken, then nobody, even the NSA nor any {\sc i-banks} will be able to get the money from this account.

\section{Deployment}
Because security is the most important thing when dealing with money, it is not safe to use a pre-released protocol for real valued exchanges. So the protocol version v0.1 is intended for testing only. People should use it, play with it as many times they want but they should not exchange goods or services against this virtual money. 
Anyone interested in playing the role of an {\sc i-bank} for testing (v0.1) can contact us.
{\sc i-banks} candidates may be accepted for testing but not for real life.

If the testing phase works well, {\sc i-banks} will receive a certificate for a period to be define to send transaction with the version 1.0 protocol, meaning this is real money.

Other modifications of the protocol in the testing phase may append; v0.2, v0.3 is necessary, before releasing the v1.0.

\section{Possible attacks}
A node receiving a transaction that will make the balance of a regular account negative shall not accept the transaction and shall not store it.

Also any transaction with a null amount value is rejected. A transaction with the sender equal to the recipient is also deleted.

Anyone can play "ping-pong" between two accounts of its own. For instance, having account A and B, one send 1\euro{} from A to B and 1\euro{} from B to A and repeat the cycle million times.
The result is that the cache system response may be slower for those account A and B, but not for other accounts. 

Any created account left more than a period of time (months) unused, with no transactions is deleted from the nodes. 

\section{Selected configuration}
The selected hashing algorithm is {\sc sha256} for all signed message.

The selected elliptic curve is NIST P-521\cite{nist}, which is stronger than the one used for {\bf Bitcoin}\cite{snakamoto}.

Messages+signatures can be sent as binary code when possible and for readable transaction, please use as possible the {\sc ascii base85}\cite{phintjens} encoding with zero padding.

Saving 8 bytes for users IDs may be considered a little large, but we must minimize the risk of collision of the same ID with different public-keys. If unfortunately that case occurs, the second user is proposed to select another couple of keys.

End user can run locally a key-generator and select the key with a base85 encoded ID she likes the most, for instance without specific characters. However hopefully, it is not possible to chose first an ID and second run in a finite time a program that will return a public-key for that ID.

\section{Anonymity}
All transactions, public-keys and certificates are public, so anyone connected to the Net can see that a given ID has bought for such amount of money to another ID (the date of the check is never a significant proof). As long as he cannot link the ID to a physical person, transaction stay anonymous. This is the same level of anonymity than for {\em Bitcoin}. 
We recommend to use several accounts, some are public (published on a web page) and some are privates (not linked to a person). Then, someone reading a transaction, does not know if this is between two different physical persons or between two accounts of the same person.

\section{Uniqueness}
It may be interesting for a citizen to select a unique digital ID and have the ability to prove to anyone that such ID is unique. In some countries, it is currently discussed the opportunity to send the same amount of money to every one every month, without any conditions, just for the fact to be in life. We can imagine how hard it is with a classical tool to check that a malicious citizen is not trying to register on two different places to get twice incomes.
This is where we need to introduce for a selected account the notion of {\bf primary account}, meaning that anybody has only one primary account at a time.

Just as for {\sc i-banks}, the {\em master authority} can deliver new certificates with a zero debt field, meaning that those institutions are not able to give money to anyone, but just able to register ID and check citizenship. Each country shall select an identification number system where a strictly unique number is given to each baby all her life. The local administration asks each citizen to see him physically with identification papers and a proof of the right national number. It is then checked on a big shared hash table on Internet that such number is not already registered.
If all tests pass, the administration will sign the id number and will associate to the account ID selected by the user. Then an unique primary account is registered.

Primary keys (public/private) are also pre-requirements for a possible future democratic voting system over the Internet. However, considering the current research results, not all requirements are satisfied to rely on Internet Voting.

\section{About Bitcoin}
One may think about the {\em Bitcoin} technology when dealing with digital money and Internet. Indeed, {\em Bitcoin} is a very interesting algorithm and curiosity for computer science. It shows how to build a digital and secure {\em Ponzi}-like pyramid. {\em Bitcoin} wants to address the difficult problem of {\bf money creation} in a pure decentralised way. First, the consequence is that the early adopters, who started in 2009, are really advantaged compared to people trying to forge coins now. This is not clearly a fare situation. Second, because it is not allowed to create money for a standard currency, say American dollar, {\em Bitcoin} is exchanged on a free market that fix a rate for the currency, and is open to strong speculation. But what is sold here ? A {\em bitcoin} is just a particular big integer with properties proving that it has been added to a block-chain started in 2009 with the random string {\em "The Times 03/Jan/2009 Chancellor on brink of second bailout for banks"} ! 

Imagine that we are creating a new $\pi$coin currency with friends. A new coin is forged each time some computer find a new decimal of $\pi$. Like {\em Bitcoin}, it is more and more difficult to create new coins, but can someone say that it owns the $n^{th}$ decimal of $\pi$ and could sell that number to someone else ? The answer is 'no' because any integer is a mathematical object belonging to everyone, means no-one. It's hopefully forbidden to claim owning a particular number, whatever are the properties of that number. The $\pi$ number at a given decimal is a much more remarkable number that a {\em bitcoin} block, but none of them should be sold as some intangible good. 
For not expert in cryptography, a big integer representing a digital signature remains a little magic, with some probably intrinsic value, and many people would be ready to pay to own one of them. But it's a pure game where each player try to convince other less experts or other gamers to enter the game in order to gain real money on behalf them. 

The scarcity of {\em Bitcoin} does not make it valuable as it has always been for gold. The name "forging" has been used to make believe it is as exhausting to extract {\em bitcoins} than extracting gold. But it is false; the object value is not the price of a human miner hard working because for {\em Bitcoin}, only machines are really "working", more exactly they are consuming some electricity to find big integers that will not help at all research in Mathematics. So {\em Bitcoin} is a nice trickery just for people who like to gamble.
The economics requirement for every day trading is not a game, banks are not lotteries. So we can no rely on a the {\em Bitcoin} family system as a payment mean for every citizen.

The following table summarize differences between {\em Bitcoin}, {\sc moi} and the traditional banking system:

\begin{center}
\begin{tabular}{ | m{3.5cm} | m{3.8cm}| m{3.8cm} | m{3.8cm} | } 
\hline
 & {\em Bitcoin} & {\sc moi} & Banks\\ 
\hline
Money type & {\em value} money & {\em debt} money & both value\&debt \\ 
\hline
Money creation issue & yes & no &yes \\ 
\hline
Centralization & none & minimal & full\\ 
\hline
Money mass  & finite & infinite & rules\\ 
\hline
Transactions history & chain required & set required & private network \\ 
\hline
Architecture & P2P & P2P & private servers \\ 
\hline
Rate with \$ & free (speculative) & fixed as identical & official \\ 
\hline
{\sc ecc} curve & P-384 & P-521 & none\\ 
\hline
Power consumption & very high & very small & manual tasks\\ 
\hline
Main purpose \$ & currencies exchange & goods\&services trading & traditional banking \\ 
\hline
\end{tabular}
\end{center}

Our proposal has some common feature with {\em Bitcoin} as both are using Elliptic Curves Cryptography ({\sc ecc}). Because we aim to offer a digital payment system for legal currencies, we do not address the problem of money creation. Thus, we did not succeed in making the all system fully decentralized. At least one {\sc i-bank} should exist and its debt bounded by a {\em master authority}. However, no entity is in charge of the big issue to create money from scratch, as the ECB or the FED in the classical system. For a really democratic and fare distribution of money to all living humans on Earth, it would be mandatory to use a {\em primary account}, insuring uniqueness, but the decision to create money such a new way would be the responsibility of governments. The {\sc moi} protocol does not require any change to be {\sc Money Over the Internet}.

\section{First Application}
The {\sc moi} protocol will be used for paying electricity consummation for electric vehicle in public areas, using the {\color{blue} \bf pluggle} system. For the end user, the same mobile phone application enable to lock/unlock chargers and to pay the owner of the charger. The application may move to a full digital generic person2person payment system after experimentation phase. 

\section{Conclusion}
This paper shows how it is simple to use {\em Internet} as a universal payment system, without {\em Bitcoin} drawbacks. This system can be called {\sc Money Over Ip} ({\sc moi}). We proposed a pre-released version of the protocol for debate and testing.
As money remains a matter of concern of any-one, not just for computer scientists, we hope that our proposal will raise discussions and new ideas with the same goal of offering to any citizen on Earth a {\bf universal, open, free} and  {\bf easy to use} payment system. Please, do not hesitate to join us for this {\sc moi} project.

\end{document}